\documentclass[11pt,twoside]{article}
\usepackage[utf8]{inputenc}
\usepackage[T1]{fontenc}
\usepackage{amsmath, amsfonts, amssymb}
\usepackage[mathscr]{eucal}
\usepackage{mathtools}
\usepackage{times}
\usepackage{bm}
\usepackage{booktabs}
\usepackage{color}
\usepackage[natural, dvipsnames, pdftex]{xcolor}
\usepackage{algorithm}
\usepackage[noend]{algpseudocode}
\usepackage[pdftex, hidelinks]{hyperref}
\usepackage[authoryear]{natbib}
\usepackage{multirow}
\usepackage{graphicx}
\graphicspath{{figure/}}

\usepackage[top=3cm, bottom=3cm, left=2.25cm, right=2.25cm]{geometry}
\usepackage[calcwidth,  sf,  big, compact]{titlesec}
\titleformat{\section}
{\normalfont\scshape \Large}{\thesection}{1em}{}
\usepackage{fancyhdr}
\usepackage{setspace}
\usepackage[amsmath,  thref,  thmmarks,  hyperref]{ntheorem}
\usepackage{cleveref}

\titleformat{\section}[block]{\large \bf }
{  {\thesection.}}{4pt}{   }
\titleformat{\subsection}[block]{\itshape}
{  {\thesubsection.}}{4pt}{   }

\def\and{\textsc{and }}

\def\pr{{\rm Pr}}

\def\calC{{\mathcal C}}

\def\upsilon{{\Upsilon}}
\def\bi{\begin{itemize}}
\def\ei{\end{itemize}}

\newcommand{\Istat}{\textsf{ISTAT}}
\newcommand{\IDLold}{\textsf{IDL 2016}}
\newcommand{\France}{\textsf{France 2019}}
\newcommand{\IDL}{\textsf{IDL}}
\newcommand{\SM}{{Appendix}}
\DeclareMathOperator{\sign}{sign}
\usepackage{wasysym}

\title{Human mortality at extreme age}

\author{L\'eo R. Belzile\thanks{HEC Montréal, Department of Decision Sciences, 3000, chemin de la Côte-Sainte-Catherine, Montréal (QC), Canada H3T 2A7 (\texttt{leo.belzile@hec.ca})}, Anthony C. Davison\thanks{EPFL-SB-MATH-STAT, Station 8, \'Ecole polytechnique f\'ed\'erale de Lausanne, 1015 Lausanne, Switzerland}, Holger Rootzén and  Dmitrii Zholud\thanks{Department of Mathematical Sciences, Chalmers and Gothenburg University, Chalmers Tvärgata 3, 41296 Göteborg, Sweden}
}
\date{}

\begin{document}

\maketitle

\begin{abstract}
We use a combination of extreme value statistics, survival analysis and computer-intensive methods to analyze the mortality of Italian and French semi-supercentenarians. After accounting for the effects of the sampling frame, we conclude that constant force of mortality beyond 108 years describes the data well and there is no evidence of differences between countries and cohorts. These findings are consistent with use of a Gompertz model and with previous analysis of the International Database on Longevity and suggest that any physical upper bound for the human lifespan is so large that it is unlikely to be approached. Power calculations suggest that it is unlikely that there could be an upper bound below 130 years.  There is no evidence of differences in survival between women and men after age 108 in the Italian data and the International Database on Longevity, but survival is lower for men in the French data.
\end{abstract}

Solid empirical understanding of human mortality at extreme age is important as one basis for research aimed at finding a cure for ageing \citep[described, e.g., in][]{vijg-campisi:2008}, and is also an element in the hotly-debated and societally-important question whether the current increase in expected lifespan in developed countries, of about three months per year since at least 1840 \citep{Oeppenetal:2002}, can continue. The limit to human lifespan, if any, also attracts considerable media attention \citep[e.g.][]{Guarino:2018,Saplakoglu:2018,Hoad:2019}.

Einmahl~et~al.~\cite{einmahl-etal:2019} analyse data on mortality in the Netherlands and conclude that ``there indeed is a finite upper limit to the lifespan'' for both men and women. Their dataset, provided by Statistics Netherlands and consisting of about 285,000 ``Dutch residents, born in the Netherlands, who died in the years 1986--2015 at a minimum age of 92 years'', had not undergone any validation procedure. As might be expected, the vast majority died before their 100th birthdays: 99.5\% lived 107 or fewer years, and 97\% died at age 101 or younger. The cohorts for analysis were taken to be the calendar years of death, and truncation of lifetimes was not taken into account. Hanayama \& Sibuya \cite{Hanayama/Sibuya:2016} also find an upper lifespan limit of 123 years for Japanese persons aged 100 or more, by fitting a generalized Pareto distribution to one-year and four-year birth cohorts, taking into account the sampling scheme. In both cases, any plateauing of mortality may be confounded with the increase in hazard between ages 100 and 105, and this would invalidate the extrapolation to extreme age. 

The validity of conclusions about mortality at extreme age depends crucially on the quality of the data on which they are based \citep{Gavrilov/Gavrilova:2019}, as age misrepresentation for the very old is common even in countries with otherwise reliable statistical data \citep{poulin2010}. Motivated by this, demographic researchers from 15 countries contribute to the International Database on Longevity (\IDL{}), which currently contains 860 individually-validated life lengths of supercentenarians who had reached age 110 or more on 28 September 2019; the data, which cover different time periods for different countries, can be obtained from \url{www.supercentenarians.org}.  For some countries the \IDL{} has recently been extended to include data on semi-supercentenarians, i.e., people who lived to an age of at least 105. As of October 2019, it contained new French data on 9612 semi-supercentenarians and 241 supercentenarians, which we call the \France{}\ data; all these supercentenarians but only some of the semi-supercentenarians were validated.   

A 2016 version of the \IDL{} (\IDLold{}) was analysed by Gampe~\citep{gampe2010} and Rootz\'en \& Zholud \citep{rootzen-zholud:2017}, the latter with extensive discussion \citep{RootzenZholud:2018}. Both papers made allowance for the sampling scheme, and in particular for the truncation of lifetimes that it entails. They concluded that there is no indication of an increase in mortality for ages above 110 years, and hence no indication of a finite upper limit to the human lifespan. Rootz\'en \& Zholud \citep{rootzen-zholud:2017} also found no differences in mortality between men and women or between populations from regions and  countries as varied as  Japan, the USA or Europe. These conclusions are striking, but the small size of the \IDL{} and the lack of balance between the subgroups limit the statistical power available to detect such differences. 
 
The Italian National Institute of Statistics (\Istat{}) has recently produced an important new database \citep{Istat2018} containing individually validated birth dates and survival times in days of all persons in Italy who were at least 105 years old at some point in the period from 1 January 2009 to 31 December 2015. Using advanced survival analysis tools, Barbi et al.~\citep{Barbi:2018} found that death rates in this dataset ``reach or closely approach a plateau after age 105'' and found a small but statistically significant cohort effect.

\begin{figure}
\centering
\includegraphics[width=0.65\linewidth]{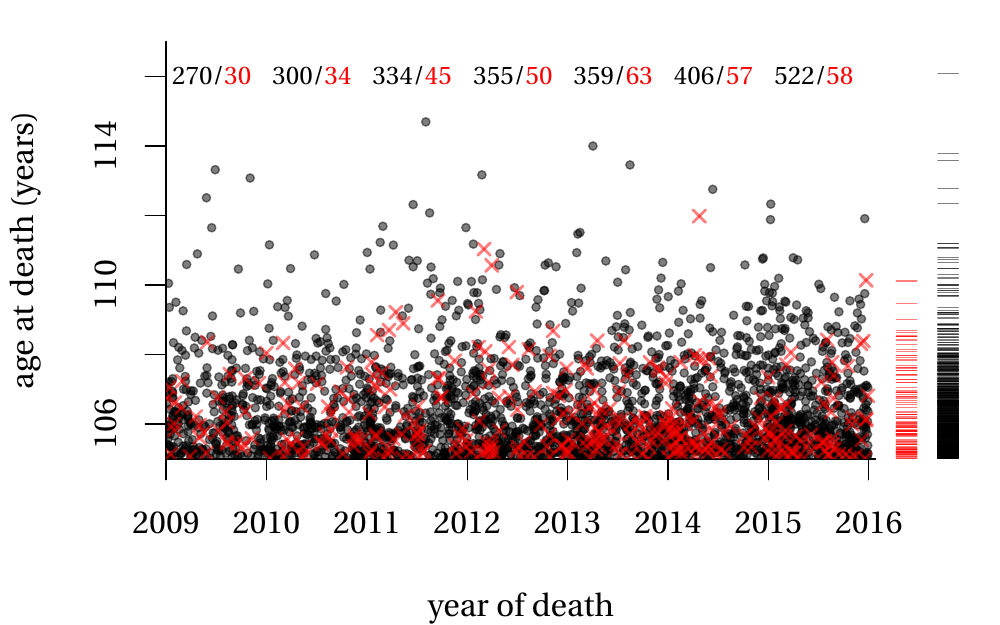}
	\caption{Lexis diagram of the \Istat{}\ data, showing age and calendar time at death for men (red crosses) and women (grey dot if only one woman, black dot if several). Censored observations are displayed in the right margin; the ticks indicate the ages of men (red) and women (black) still alive on 31 December 2015. The sampling frame includes only persons aged at least 105 years and alive on 1 January 2009, or whose 105th birthday occurred before 1 January 2016. The death counts for each calendar year are given at the top of the graph for men (red) and women (black).}
	\label{fig:Istatdata}
\end{figure}

Data analysis must take into account the sampling scheme underlying such data.  The \Istat{} lifetimes are left-truncated because only individuals who attain an age of at least 105 years  during the sampling period are included, and they are right-censored because the date of death of persons still alive in 2016 is unknown; see \Cref{fig:Istatdata}. The right-censored lifetimes are shown by the tick marks at the  right side of the panel, and include the oldest person; ignoring  either the truncation or the censoring could lead to incorrect conclusions.  The \France{} lifetimes are left- and right-truncated: only individuals who are observed to die during the sampling period appear in the dataset.  The statistical consequences are discussed in the \SM~\ref{subsection:SIB}.

 In our analysis we take the sampling frame into account and pinpoint the age, if any, at which mortality attains a plateau, and disentangle the effects of age and of birth cohort. We also compare mortality in the \Istat{}, \France{}\, and \IDLold{} databases, and between men and women.
      
We use the generalized Pareto distribution from extreme value statistics in the main analysis, supplemented by fits of the Gompertz distribution, which is standard in demography.  We first outline our main results and conclusions; the \SM\ gives a more detailed description of our methods.

\section*{Results for ISTAT data}%
\label{sect:ISTATresults}

Lifetimes beyond 105 years are highly unusual and the application of extreme value models \citep{dehaanferreira2006} is warranted. We use the generalized Pareto distribution,
\begin{equation}
\label{eq:GP}
F(x) = \begin{cases}
 1-(1+{\gamma}x/{\sigma})_+^{-1/{\gamma}}, & x \geq 0, \gamma \neq 0, \\
 1-e^{-{x}/{\sigma}}, & x \geq 0, \gamma = 0,
 \end{cases}
\end{equation}
to model $x$, the excess lifetime above $u$ years. In~\eqref{eq:GP}, $a_+=\max(a,0)$ and $\sigma > 0$ and $\gamma \in \mathbb{R}$ are scale and shape parameters. For negative shape parameter $\gamma$ the distribution has a finite upper endpoint at $-\sigma/\gamma$, whereas $\gamma\geq 0$ yields an infinite upper endpoint.

The corresponding hazard function, often called the ``force of mortality'' in demography, is the density evaluated at excess age $x$, conditional on survival to then, i.e., 
\begin{equation}
\label{eq:GPhazard}
h(x)=\dfrac{f(x)}{1-F(x)} = \dfrac{1}{(\sigma + \gamma x)_+}, \quad x\geq 0, 
\end{equation}
where $f(x)=\mathrm{d}F(x)/\mathrm{d}x$ is the generalized Pareto density function. If $\gamma <0$,  the hazard function tends to infinity at the finite upper limit for exceedances. When $\gamma = 0$,  $F$ is exponential and the hazard function is constant, meaning that the likelihood that a living individual dies does not depend on age beyond the threshold. In this case, mortality can be said to have plateaued at age $u$.  

\begin{figure}%[tbhp]
\centering
\includegraphics[width=0.65\linewidth]{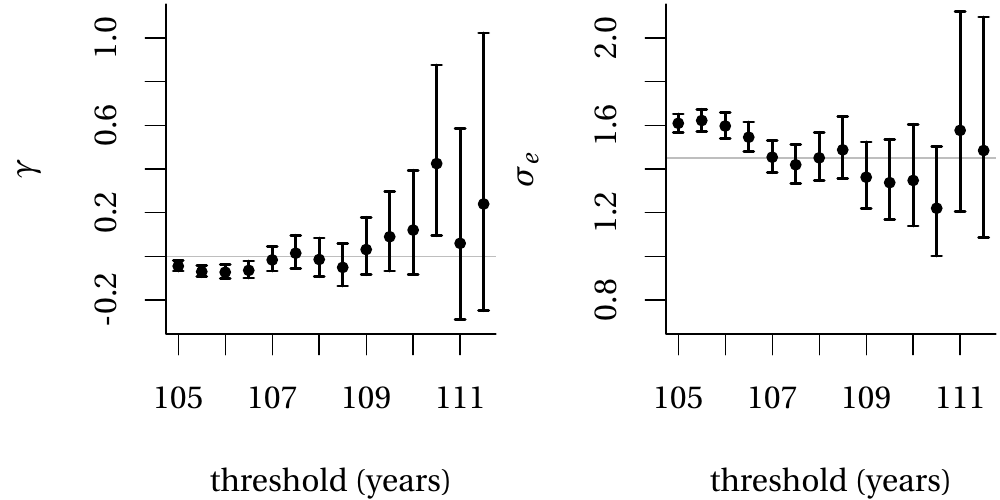}
\includegraphics[width=0.65\linewidth]{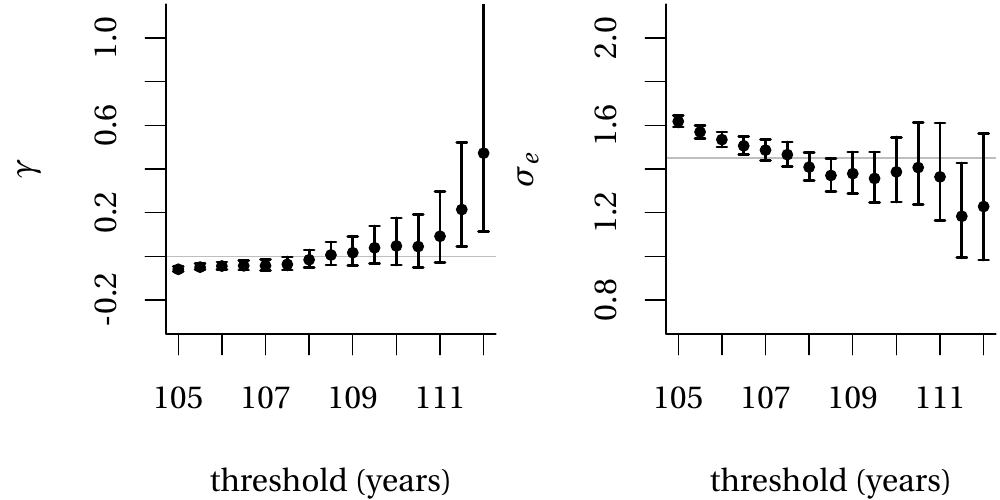}
\caption{Parameter stability plots for the \Istat{} data (top) and for the \France{}  data (bottom), showing the shape $\gamma$ of the generalized Pareto distribution (left) and the scale $\sigma_e$ of the exponential distribution (right) based on lifetimes that exceed the age threshold on the $x$-axis. The plots give maximum likelihood estimates with 95\% confidence intervals derived using a likelihood ratio statistic. The horizontal lines in the right-hand panels correspond to the estimated scale for excess lifetimes over 108 years for the \Istat{} data.}
\label{fig:parameterstability}
\end{figure}

The choice of a threshold $u$ such that Eq.~[\ref{eq:GP}] models exceedances appropriately is a basic problem in extreme value statistics and is surveyed by  Scarrott \& MacDonald~\cite{scarrott/macdonald:2012}. If $u$ is high enough for Eq.~[\ref{eq:GP}] to provide an adequate approximation to the distribution of exceedances, then the shape parameter $\gamma$ is approximately unchanged if a higher threshold $u'$ is used, and the scale parameters for $u$ and $u'$ have a known relationship, so a simple and commonly-used approach to the choice of threshold is to plot the parameters of the fitted distributions for a range of thresholds \citep{davison+s:1990} and to use the lowest threshold above which parameter estimates stabilise. This choice balances the extrapolation bias arising if the threshold is too low with the increased variance incurred when taking $u$ too high to retain an adequate number of observations.

The upper left-hand panel of Figure~\ref{fig:parameterstability} shows that for age thresholds close to 105 years the estimated shape parameters for excess life lengths are negative, with 95\% confidence intervals barely touching zero,  but there is no systematic indication of non-zero shape above 107 years.   The upper right-hand  panel displays the estimated scale parameter of the exponential model fitted to life lengths exceeding the threshold. The scale parameters decrease for ages 105--107 but show no indication of change after age 107, where the scale parameter estimate is 1.45. Parameter stability plots suggest an exponential model and hence a constant hazard after age 107 or so, where a mortality plateau seems to be attained.

The upper part of \Cref{tab:ISTAT-MLE} shows results from fitting Eq.~[\ref{eq:GP}] and the exponential distribution to the \Istat{}\  for a range of thresholds.  The exponential model provides an adequate fit to the exceedances over a threshold at 108 years, above which  the hypothesis that $\gamma=0$, i.e., the exponential model is an adequate model simplification, is not rejected.

% latex table generated in R 3.6.1 by xtable 1.8-4 package
% Mon Nov 18 23:54:21 2019
\begin{table*}[t]
\centering

\begingroup\footnotesize
\begin{tabular}{lrrrrrrrr}
  \toprule
%  \multicolumn{8}{c}{\Istat{} data}\\
%   \midrule
\Istat{} &  threshold & $105$  & $106$  & $107$  & $108$  & $109$  & $110$  & $111$\\
 \midrule
&  $n_u$ & $3836$ & $1874$ & $947$ & $415$ & $198$ & $88$ & $34$ \\ 
&  $\sigma$ & $1.67\; (0.04)$ & $1.7\; (0.06)$ & $1.47\; (0.08)$ & $1.47\; (0.11)$ & $1.33\; (0.15)$ & $1.22\; (0.23)$ & $1.5\; (0.47)$ \\ 
&    $\gamma$ & $-0.05\; (0.02)$ & $-0.07\; (0.03)$ & $-0.02\; (0.04)$ & $-0.01\; (0.06)$ & $0.03\; (0.09)$ & $0.12\; (0.17)$ & $0.06\; (0.30)$ \\ 
&    $\sigma_e$ & $1.61\; (0.03)$ & $1.6\; (0.04)$ & $1.45\; (0.05)$ & $1.45\; (0.08)$ & $1.36\; (0.11)$ & $1.35\; (0.17)$ & $1.58\; (0.32)$ \\ 
&    $p$-value & $0.04$ & $0.01$ & $0.70$ & $0.82$ & $0.74$ & $0.44$ & $0.84$ \\ 
&    $p_\infty$ & $0.02$ & $0.01$ & $0.35$ & $0.41$ & $0.63$ & $0.78$ & $0.58$ \\ 
   \bottomrule
%\end{tabular}
%\endgroup
%\begin{table*}[ht]
%\centering
%\caption{Estimates (standard errors) of scale  and shape parameters ($\sigma$, $\gamma$) for the generalized Pareto distribution and of the scale parameter ($\sigma_e$) for the exponential model for the French \IDL{} data as a function of threshold, with number of threshold exceedances ($n_u$), $p$-value for the likelihood ratio test of $\gamma=0$ and probability under the fitted model that the endpoint is infinite, corresponding to $\Pr(\gamma>0)$.} 
%\begingroup\footnotesize
%\begin{tabular}{lrrrrrrrr}
\toprule
\France{} &   threshold & $105$  & $106$  & $107$  & $108$  & $109$  & $110$  & $111$\\
 \midrule
&  $n_u$ & $9835$ & $5034$ & $2472$ & $1210$ & $550$ & $240$ & $106$ \\ 
&    $\sigma$ & $1.69\; (0.02)$ & $1.59\; (0.03)$ & $1.54\; (0.04)$ & $1.43\; (0.06)$ & $1.36\; (0.08)$ & $1.34\; (0.13)$ & $1.26\; (0.18)$ \\ 
&    $\gamma$ & $-0.06\; (0.01)$ & $-0.04\; (0.01)$ & $-0.04\; (0.02)$ & $-0.02\; (0.03)$ & $0.02\; (0.05)$ & $0.05\; (0.07)$ & $0.09\; (0.11)$ \\ 
&    $\sigma_e$ & $1.62\; (0.02)$ & $1.53\; (0.03)$ & $1.49\; (0.03)$ & $1.41\; (0.05)$ & $1.38\; (0.07)$ & $1.39\; (0.11)$ & $1.36\; (0.16)$ \\ 
&    $p$-value & $4 \times 10^{-7}$ & $0.01$ & $0.05$ & $0.60$ & $0.72$ & $0.48$ & $0.32$ \\
&   $p_\infty$  & $2 \times 10^{-7}$ & $2 \times 10^{-3}$ & $0.02$ & $0.30$ & $0.64$ & $0.76$ & $0.84$ \\ 
   \bottomrule
\end{tabular}
\caption{Estimates (standard errors) of scale  and shape parameters ($\sigma$, $\gamma$) for the generalized Pareto distribution and of the scale parameter ($\sigma_e$) for the exponential model for the \Istat{} and \France{} datasets as a function of threshold, with number of threshold exceedances ($n_u$), $p$-value for the likelihood ratio test of $\gamma=0$ and probability that $\gamma \geq 0$ based on the profile likelihood ratio test under the generalized Pareto model ($p_\infty$).} 
\label{tab:ISTAT-MLE}
\endgroup
%\end{table*}
\end{table*}

%In short: parameter stability plots point to the choice of $u=10?$ as threshold for generalized Pareto modelling of excess life lengths distribution and give no indication of deviations from an exponential distribution above $u$; cumulative hazard and hazard estimates point to a constant hazard above $u$; and Bayesian 90\% credible intervals for a possible limit for human life lengths based on excess life lengths above $u$ or more all include $\infty$. Further, Figure~\ref{fig:fit} shows a good fit of the exponential distribution to the \Istat{} data for persons who lived to age $u$. This all indicates that increases in mortality at younger ages stopped around age 107 in the \Istat{} data.

The estimated scale parameter obtained by fitting an exponential distribution to the \Istat{}\ data for people older than 108 is 1.45 (years)  with 95\% confidence interval $(1.29, 1.61)$.  Hence the hazard is estimated to be 0.69 (1/years) with 95\% confidence interval $(0.62, 0.77)$; above 108 years the estimated probability of surviving at least one more year at any given age is 0.5 with 95\%  confidence interval $(0.46, 0.54)$.

We investigated birth cohort effects, but found none; see \SM~\ref{subsect:cohorteffect}.

% This claim is further supported by a Bayesian analysis reported in \SM\ \ref{subsect:bayes}. 

\section*{Results for \France{} data}%
\label{sect:Frenchresults}

% \begin{figure}[t!]
% 	\centering
% 	
% 	\caption{Parameter stability plots for the shape $\gamma$ of the generalized Pareto distribution (left) and for the scale $\sigma_e$ of the exponential distribution (right), for lifetimes that exceed the age threshold on the $x$-axis for the \France{}  data. The plots give maximum likelihood estimates with (profile) likelihood-based 95\% confidence intervals. }
% 	\label{fig:frenchstab}
% \end{figure}

\begin{table*}[t!]
	\centering
		
	\begin{tabular}{l  rl l rl l rl}
		\toprule
 & \multicolumn{2}{c}{\Istat{}}& & \multicolumn{2}{c}{\France{}}&&\multicolumn{2}{c}{\IDLold{}} \\
 \cline{2-3}\cline{5-6}\cline{8-9}\\
		 & \multicolumn{1}{c}{$n$} &\multicolumn{1}{c}{$\sigma_e$ (95\% CI)}& &  \multicolumn{1}{c}{$n$} & \multicolumn{1}{c}{$\sigma_e$ (95\% CI)} & &\multicolumn{1}{c}{$n$} & \multicolumn{1}{c}{$\sigma_e$  (95\% CI)}    \\
	 \midrule
		{\footnotesize women }& $375$ &$1.45~(1.23, 1.62)$ && $1116$ & $1.46~(1.36, 1.56)$&& $507$ & $1.39~(1.25, 1.54)$\\
		{\footnotesize men } &$40$ &$1.41~(0.86, 1.98)$&& $94$  & $0.90~(0.70, 1.11)$ && $59$& $1.68~(1.16, 2.20)$  \\
 {\footnotesize All}& $415$ &$1.45~(1.29, 1.61)$ && $1210$ & $1.41~(1.32, 1.51)$ && $566$ & $1.42~(1.28, 1.56)$ \\
 \bottomrule
		\end{tabular}
		\caption{Estimates of the scale, $\sigma_e$, of the exponential distribution, with 95\% confidence intervals (CI). This distribution is fitted to exceedances of 108 years in the \Istat{}\ and \France{} data and of 110 years in the \IDLold{}\ data analysed in \cite{rootzen-zholud:2017}. }
\label{table:women-men}
		\end{table*}		

Estimation for the \France{} data was done as described in Rootz\'en \& Zholud \citep{rootzen-zholud:2017}, taking into account the left- and right truncation of the lifetimes. Both the parameter stability plots in the lower panels of \Cref{fig:parameterstability} and the results given in the lower part of Table~\ref{tab:ISTAT-MLE} indicate that the exponential model is adequate above 108 years. For persons older than 108 the exponential scale parameter is estimated to be 1.41 (years)  with 95\% confidence interval $(1.32, 1.51)$, the hazard is estimated to be 0.71 (years$^{-1}$) with 95\% confidence interval $(0.66, 0.76)$ and the estimated probability of surviving at least one more year is  0.49 with 95\%  confidence interval $(0.47, 0.52)$.
 
Table~\ref{table:women-men} shows that estimates of the scale parameter for the exponential distribution for women and men for the  \France{} data differ. If men are excluded, then  the estimated scale parameter increases from 1.41 to 1.46 years, and if the oldest woman, Jeanne Calment, is also excluded, the estimate for women drops to 1.44 years. Similarly to the \Istat{} data, survival for ages 105--107 was lower in earlier cohorts. 

\section*{Power}

\section*{Power}

Our analysis above suggests that there is no upper limit to human lifetimes and that constant hazard adequately models excess liftime if one considers only those persons whose lifetimes exceed 108 years: there is no evidence that the force of mortality above this age is other than constant.  One might wonder whether increasing force of mortality would be detectable, however, as the number of persons attaining such ages is relatively small.  To assess this we performed a simulation study described in the~\SM~\ref{subsect:power}, mimicking the sampling schemes of the \Istat{}, \France{} and \IDLold{} datasets as closely as possible and generating samples from the generalized Pareto distribution with $-0.25\leq \gamma \leq 0$. To remove overlap between the last two datasets, we dropped France from   \IDLold{}.

\begin{figure*}%[tbhp]
\centering
\includegraphics[width=0.95\linewidth]{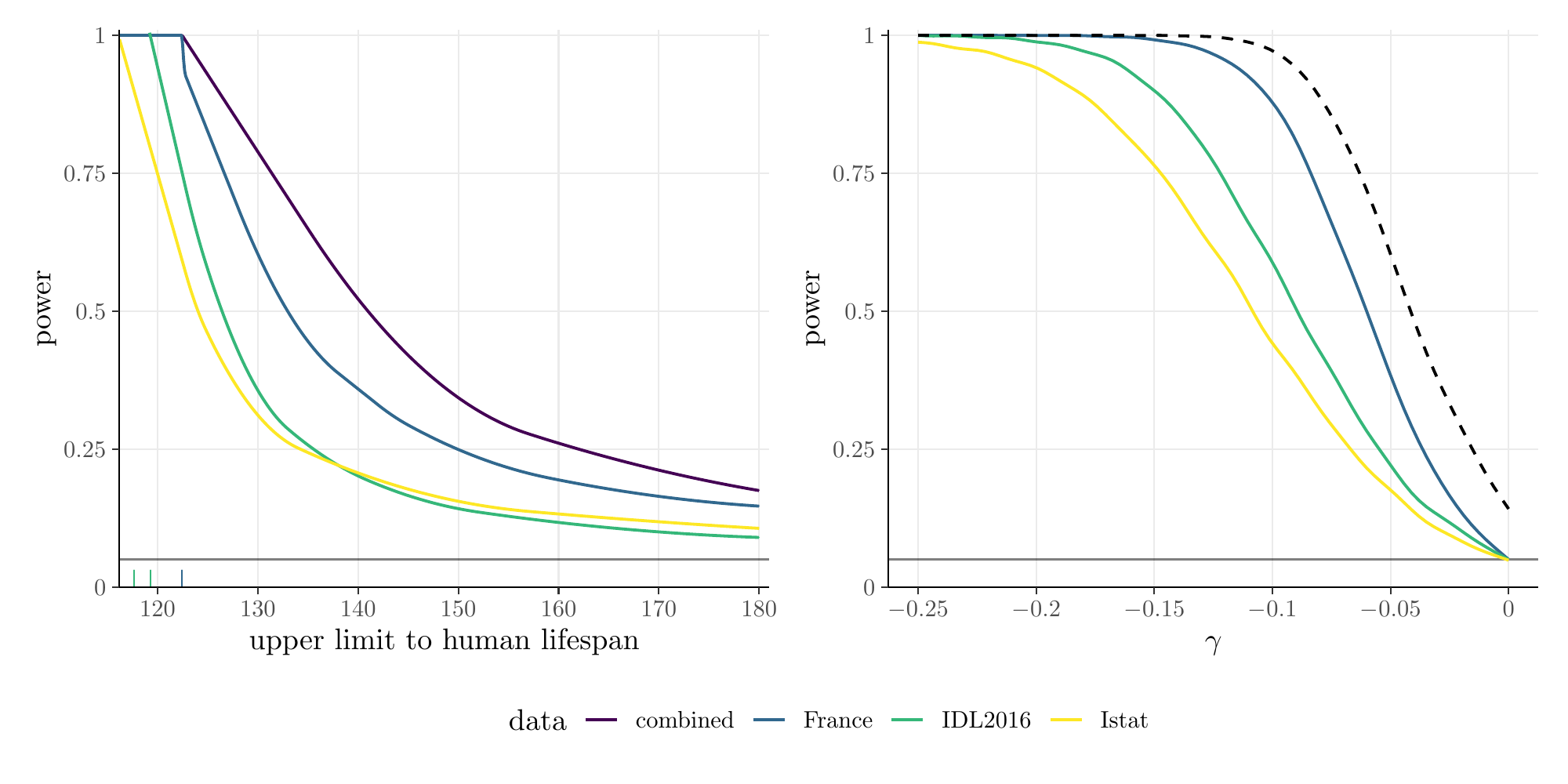}
\caption{Power functions based on the \IDLold{} (excluding French records), \France{} and \Istat{} databases and combined datasets, with rugs showing the lifetimes above 115. Left: power for the alternative of a finite endpoint $\iota$ against the null hypothesis of infinite lifetime based on the likelihood ratio statistic. The endpoint cannot be lower than the largest observations in each database. Right: power of the Wald statistic for the  null hypothesis $ \gamma = 0$ against the one-sided alternative $ \gamma < 0$ as a function of $\gamma$; the dashed line represents the power obtained by rejecting the null of exponentiality when any of the three one-sided test rejects. The curves are obtained by conditioning on the birthdates and left-truncated values in the databases, then simulating generalized Pareto data whose parameters are the partial maximum likelihood estimates $(\widehat{\sigma}_\gamma, \gamma)$. The simulated records are censored if they fall outside the sampling frame for the \Istat{} data and are simulated from a doubly truncated generalized Pareto distribution for \IDLold{} and \France{}. See \SM~\ref{subsect:power} for more details.}
\label{fig:powerendpoint}
\end{figure*}

Any biological limit to their lifespan should be common for all humans, whereas differences in mortality rates certainly arise due to social and medical environments and can be accommodated by letting hazards vary by factors such as country or sex. With the overlap dropped we can treat the datasets as independent and compute the power for a combined likelihood ratio test of $\gamma=0$ (infinite lifetime) against alternatives with $\gamma<0$ (finite lifetime).  For concreteness of interpretation we express the results in terms of the implied upper limit of lifetime $\iota=u-\sigma/\gamma$. The left-hand panel of \Cref{fig:powerendpoint} shows the power curves for the three datasets individually and pooled.  The power of the likelihood ratio test for the alternatives $\iota \in \{125, 130, 135\}$ years, for example, is $0.46/0.32/0.24$ for the \Istat{} data above 108, $0.82/0.61/0.46$ for the \France{} data above 108, and $0.62/0.40/0.29$ for the \IDLold{} data above 110. The power for $\iota=125/130/135$ years based on all three datasets is $0.96/0.80/0.64$, so it appears to be rather unlikely that an upper limit to the human lifespan, if there is such a limit, is below 130 years or so.

Similar calculations give the power for testing the null hypothesis $\gamma=0$ against alternatives $\gamma<0$. Forcing all datasets to have the same shape parameter would allow them to have different endpoints so we reject the overall null hypothesis if we reject the exponential hypothesis any of the three datasets. The power of this procedure is shown with a dashed black line in \Cref{fig:powerendpoint}. The resulting combined power exceeds $0.8$ for $\gamma < -0.07$ and equals $0.97$ for the alternative $\gamma=-0.1$, giving strong evidence against a sharp increase in the hazard function after 108 years. 

% If we treat the datasets as independent (removing the French from \IDLold{}), we can compute the power for a combined likelihood ratio test of $\gamma=0$ against alternatives $\gamma<0$.
% For example, the power of the likelihood ratio test for the alternatives $\gamma \in \{-0.15,-0.1, -0.05\}$ is $0.86/0.57/0.27$ for  the \Istat{} data above 108, $0.996/0.91/0.46$ for the \France{} data above 108 and $0.94/0.69/0.31$ for the \IDLold{} data (above 110), yielding combined power of $1-(1-0.57) (1-0.91)(1-0.69)=0.988$ for testing the alternative $\gamma=-0.1$ based on all three datasets.   Similar computations show that the joint power of the one-sided test exceeds 80\% for $\gamma < -0.06$.  

%  \blue{This high power effectively rules out the possibility of an upper limit to the human lifespan below 131 years or so.}
% \footnote{\blue{Based on the first column of Table 1, computations shown in the .tex file.}}

% Combining the estimates for threshold 105 (because the estimates of gamma are close to -0.06
%
%> gamma <- -(0.05/0.02^2 + 0.06/0.01^2)/(1/0.02^2+1/0.01^2)
%> gamma
%[1] -0.058
%> sigma <- (1.67/0.04^2 + 1.69/0.02^2)/(1/0.04^2 + 1/0.02^2)
%> sigma
%[1] 1.686
%> 105-sigma/gamma
%[1] 134.069
%
% and for threshold 108 (but the estimates of gamma are close to -0.02, for which the power is probably low
%
%> gamma <- -(0.01/0.06^2 + 0.02/0.03^2)/(1/0.06^2+1/0.03^2)
%> gamma
%[1] -0.018
%> sigma <- (1.47/0.11^2 + 1.43/0.06^2)/(1/0.11^2 + 1/0.06^2)
%> sigma
%[1] 1.439172
%> 108-sigma/gamma
%[1] 187.954

\section*{Gompertz model}

The hazard function of the generalized Pareto distribution cannot model situations in which the hazard increases to infinity, but the upper limit to lifetimes is infinite.  This possibility is encompassed by the Gompertz distribution \citep{Gompertz:1825}, which has long been used for modelling lifetimes and often provides a good fit to data at lower ages \citep[e.g.,][]{Thatcher:1999}.  When the Gompertz model is expressed in the form
\begin{align*}
F(x) = 1 - \exp\left\{-(e^{\beta x/\sigma }-1)/\beta\right\}, \quad x>0,\quad \sigma, \beta>0,             
\end{align*}
 $\sigma$ is a scale parameter with the dimensions of $x$, and the dimensionless parameter $\beta$  controls the shape of the distribution.  Letting $\beta\to 0$ yields the exponential distribution with mean $\sigma$; small values of $\beta$ correspond to small departures from the exponential model.  The fact that $\beta$ cannot be negative affects statistical comparison of the Gompertz and exponential models; see~\SM~\ref{subsect:gompertz}.

The Gompertz distribution has infinite upper limit to its support, so it cannot be used to assess whether there is a finite upper limit to the human lifespan.  Its hazard function, $\sigma^{-1}\exp(\beta x/\sigma)$, is finite but increasing for all $x$ ($\beta>0$) or constant ($\beta=0$).   The limiting distribution for threshold exceedances of Gompertz variables is exponential, and this limit is attained rather rapidly, so a good fit of the Gompertz distribution for lower $x$ would be compatible with good fit of the exponential distribution for threshold exceedances at higher values of $x$. 

Computations summarised in~\SM~\ref{subsect:gompertz} show that the exponential model, and hence also the Gompertz model with very small $\beta$, give equally good fits to the Italian and the French datasets above age 107, and that the Gompertz and generalised Pareto models fit equally well above age 105.

\section*{Conclusions}%
\label{sect:concluskons}
None of the  analyses of the \Istat{}, \IDLold{} or \France{} data, for women and men separately or combined, indicates any deviation from  exponentially distributed residual lifetimes, or equivalently from constant force of mortality, beyond 108 years.

Table~\ref{table:women-men} shows no differences between survival after age 108 in the \Istat{} data and survival after age 110 in the \IDLold{}  data for women, for men, or for women and men combined, so we merged these estimates by taking a weighted average with weights inversely proportional to the estimated variances. The resulting estimates also show no significant differences in survival between men and women, and we conclude that survival times in years after age 108 in the \Istat{} data and after age 110 in the \IDLold{} data are exponentially distributed with estimated scale parameter $1.43$ and 95\% confidence interval $(1.33, 1.52)$.  The corresponding estimated probability of surviving one more year is $0.5$ with 95\% confidence interval $(0.47, 0.52)$. 

There was no indication of differences in survival for women or the whole of the  \France{} data and in the combined  \Istat{} and \IDLold{} data, but survival for men  was lower in the \France{} data. A weighted average of the estimates for the \Istat{} data, the \France{} data and the \IDLold{} data with France removed gives an exponential scale parameter estimate of $1.42$ years with 95\% confidence interval $(1.34, 1.49)$, and estimated probability $0.49 (0.47, 0.51)$ of surviving one  more year.   

Deleting the men from the \France{} data or dropping Jeanne Calment changes estimates and confidence intervals by at most one unit in the second decimal.

There is high power for detection of an upper limit to the human lifespan up to around 130 years, based on fits of the generalized Pareto model to the three datatbases.  Moreover there is no evidence that the Gompertz model, with increasing hazard, fits better than the exponential model, constant hazard, above 108 years.

\section*{Discussion}
\label{sec:discussion}

The results of the analysis of the newly-available \Istat{}\ data agree strikingly well with those for the  \IDL{}\ supercentenarian database and for the women in the \France{}\  data.  Once the effects of the sampling frame are taken into account by allowing for truncation and censoring of the ages at death, a model with constant hazard after age 108 fits all three datasets well; it corresponds to a constant probability of 0.49 that a living person will survive for one further year, with 95\% confidence interval (0.48, 0.51).  The power calculations make it implausible that there is an upper limit to the human lifespan of 130 years or below.

Although many fewer men than women reach high ages,  no difference in survival between the sexes is discernible in the \Istat{} and the \IDLold{} data. Survival of men after age 108 is lower in the \France{} data, but it seems unlikely that this reflects a real difference between France and Italy and between France and the other countries in the \IDL{}.  It seems more plausible that this effect is due to some form of age bias or is a false positive caused by multiple testing. 

If the \Istat{} and \France{} data are split by birth cohort, then we find roughly constant mortality from age 105 for those born before the end of 1905, whereas those born in 1906 and later have lower mortality for ages 105--107; this explains the cohort effects detected by \cite{Barbi:2018}. Possibly the mortality plateau is reached later for later cohorts.  The plausibility of this hypothesis could be weighed if further high-quality data become available.

\appendix
\section{Supplementary material}
\label{sec:methods}
This section summarises the methods used to produce figures and perform our inferences, and adds goodness of fit diagnostics and hazard estimates.
%In this section we discuss how the truncation and censoring in the \Istat{}\ data were taken into account, describe the methods used for non-parametric hazard estimation and Bayesian analysis, and  investigate the robustness of our conclusions. 
%Some analyses were performed using the publicly-available Matlab toolbox \href{https://github.com/OGCJN/Rejoinder-to-discussion-of-the-paper-Human-life-is-unlimited---but-short}{LATool}, which makes alternative analyses straightforward. 
%\b{Add something about R and reproducibility? All graphs and all but one table produced using R. Our code must be provided anyway.}

\subsection{Reproducibility}%

The \Istat{} data can be obtained from the Italian National Institute of Statistics by registering at the Contact Center (https://contact.istat.it) and mentioning the Semisupercentenarian Survey and Marco Marsili as contact person.  

\href{https://github.com/OGCJN/Rejoinder-to-discussion-of-the-paper-Human-life-is-unlimited---but-short}{LATool} is a \textsf{MATLAB} toolbox for life length analysis that makes alternative analyses possible. It consists of three files that are available from \texttt{https://doi.org/10.1007/s10687-017-0305-5}, and a data file, \texttt{data.mat}, which can be downloaded, together with the full IDL database, by registering at \href{https://www.supercentenarians.org}{\texttt{www.supercentenarians.org}}. 

%The toolbox may also be obtained from the third or fourth authors, \texttt{hrootzen@chalmers.se} or  \texttt{dzholud@chalmers.se}.  

\textsf{R} \cite{R2020} was used for many of the analyses and to generate all the figures as well as \Cref{tab:ISTAT-MLE,tab:gompertz}. Standalone code to reproduce the analyses is \href{https://github.com/lbelzile/supercentenarian}{provided online}.

\subsection{Truncation and censoring}%
\label{subsection:SIB}
\Cref{fig:Istatdata} shows that many persons in the \Istat{}\ dataset were alive on 31 December 2016, when sampling finished, so their lifetimes must be treated as right-censored.   Lifetimes above 105 years on 1 January 2009 are left-truncated and individuals not aged at least 105 from 1 January 2009 to 31 December 2016 are not included. Both censoring and truncation must be handled correctly to avoid biased inferences; the effect of the truncation is that inferences must be conditioned on the event that an individual appears in the database. Below we outline how this is achieved; see \cite{rootzen-zholud:2017} and \cite{Davison:2018} for more details.

Consider a sampling interval $\calC=(b,e)$ of calendar time during which individuals aged over a threshold of $u$ years were observed.  Let $x = \mathrm{age} - u$ denote the excess age of an individual who dies aged older than $u$, having reached age $u$ at calendar time $t$. Assume that the excess ages $x$ are independent with cumulative distribution function $F(x;\boldsymbol{\theta})$, probability density function $f(x; \boldsymbol{\theta})$, and survival function $S(x, \boldsymbol{\theta})=1-F(x, \boldsymbol{\theta})$, where $\boldsymbol{\theta}$ is a vector of parameters to be estimated. 
%In this paper, $F$  denotes either a generalized Pareto or an exponential distribution.

The likelihood contribution for someone who died in $\calC$ is then
\begin{equation*}\label{eq:likelihood-daed}
 \frac{f(x)}{S\{(b-t)_+\}},\quad x>0, 
\end{equation*}
%where $(y_i)_+ = \max\{y_i, 0\}$, 
whereas that for someone who is known to be alive at the end of $\calC$, and thus whose lifetime is censored,  is
\begin{equation*}\label{eq:likelihood-alive}
\frac{S(e-t)}{S\{(b-t)_+\}}, \quad t<e.
\end{equation*}
The likelihood function $ L(\boldsymbol{\theta})$  is the product of the likelihood contributions from all individuals included in the data under study. Estimates for the generalized Pareto and Gompertz distributions were found by numerical maximization of the log likelihood, with standard errors obtained from the inverse observed information matrix. Explicit expressions exist for the maximum likelihood estimator of the exponential distribution scale parameter and its standard error,
\begin{equation}
    \label{eq:exp-est}
  \widehat{\sigma}_e = \frac{\sum_i\{x_i-(b-t_i)_+\}}{\#\mbox{deaths}},
 \quad 
  \mathrm{se}(\widehat{\sigma}_e) = \frac{\widehat{\sigma}_e}{\sqrt{\#\mbox{deaths}}}.
\end{equation}
%\begin{align}\label{eq:exp-est}
%  \widehat{\sigma} &= %\frac{\sum(x_i-(b-t_i)_+)}{\#\mbox{deaths}},
%  \\ %\label{eq:exp-sd}
%  \mathrm{se}(\widehat{\sigma}) &= \frac{\widehat{\sigma}}{\sqrt{\#\mbox{deaths}}}.\nonumber
%\end{align}
The \IDL{} and \France{} data were left- and right-truncated, and this was taken into account in our analysis;
the likelihood contribution for these individuals is 
\begin{equation}
 \frac{f(x_i)}{F(e-t_i) - F\{(b-t_i)_+\}},\qquad b-t_i \leq x_i \leq e-t_i. 
\end{equation}

Inappropriate analysis can lead to misleading results: for example, fitting an exponential distribution to the \Istat{}\ individuals who survive beyond age 107 without accounting for truncation or censoring gives the estimate $\widehat{\sigma}_e = 1.25$ with 95\% confidence interval $(1.17, 1.33)$, to be compared with $\widehat{\sigma}_e = 1.45$ with 95\% confidence interval $(1.35, 1.56)$ once the truncation and censoring are accounted for; these confidence intervals do not overlap.

\subsection{\Istat{} cohort effects}
\label{subsect:cohorteffect}

\begin{figure}[t!]
\centering 
\includegraphics[width=0.55\linewidth]{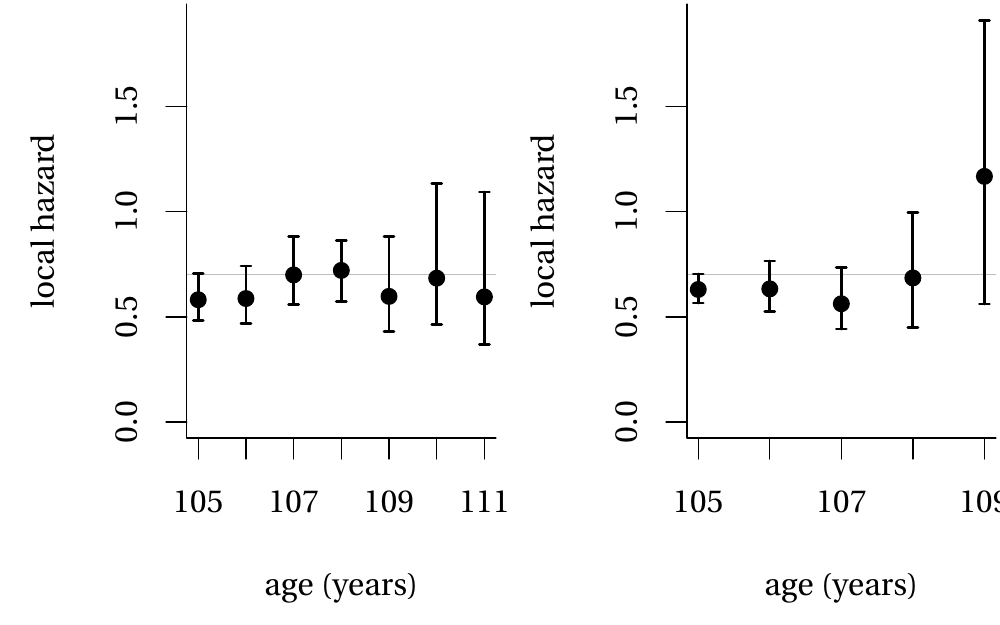}
\caption{Local hazard estimates with 95\% pointwise confidence intervals for \Istat{} birth cohorts 1886-1905 (left) and 1906--1910 (right), with horizontal lines at 0.7. The rightmost point includes all survivors beyond 111 years (respectively 109 years).}
\label{fig:hazard-birth}
\end{figure}

The local hazard estimates in \Cref{fig:hazard-birth} are obtained by splitting the likelihood contribution of individuals into yearly blocks, using disjoint intervals with $(b,e)=(a,a+1)$ years to avoid using the same individuals several times. For the highest interval we set $e=\infty$ and include all individuals who survived into that interval. 

\begin{figure}[t!]
	\centering
\includegraphics[width=0.6\linewidth]{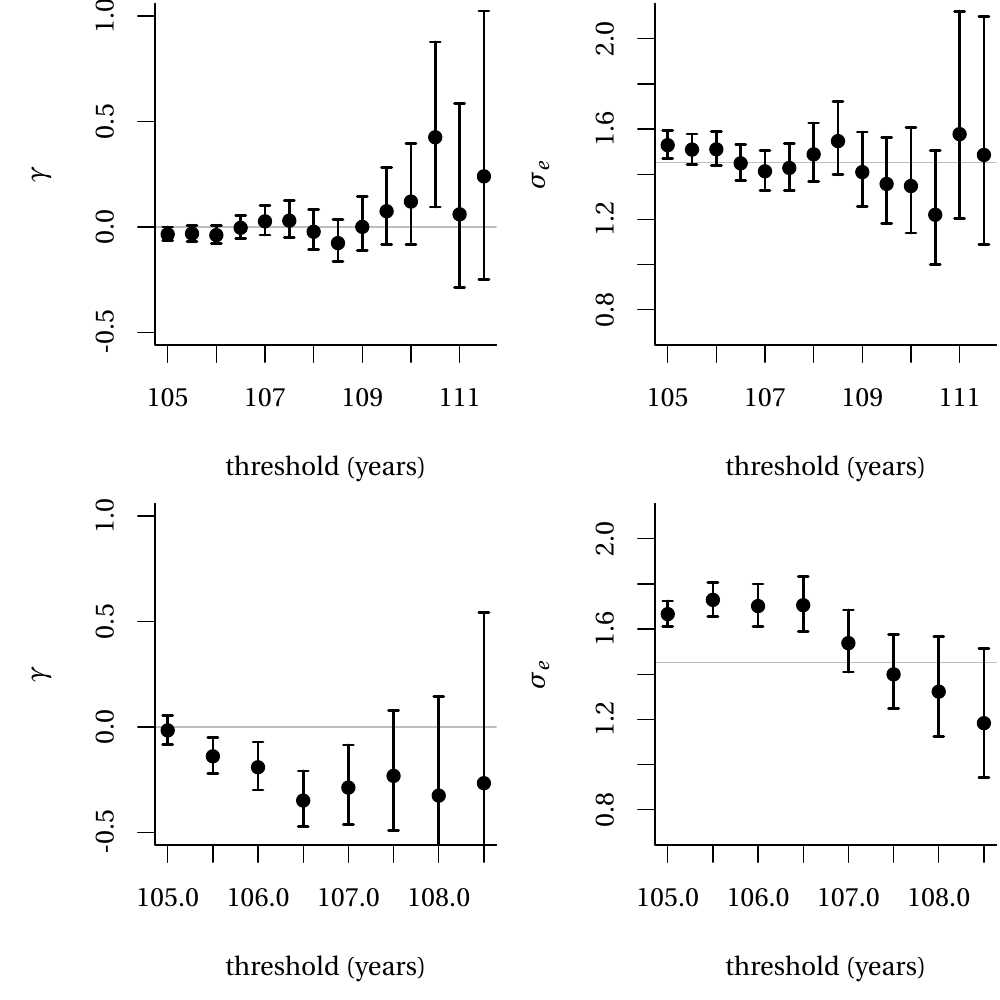}
\caption{Parameter stability plots for the \Istat{} birth cohorts 1896--1905 (top) and 1906--1910 (bottom), for the parameters $\gamma$ of the generalized Pareto distribution (left) and for the scale $\sigma_e$ of the exponential distribution (right)  obtained from exceedances of the age threshold on the $x$-axis. The plots give maximum likelihood estimates with (profile)  likelihood 95\% confidence intervals.  The horizontal line on the right panels corresponds to the estimated scale for exceedances above 108 for the full \Istat{} dataset.}
	\label{fig:parameterstability-birth}
\end{figure}

The parameter stability plots in \Cref{fig:parameterstability-birth} and the estimated hazard plots in \Cref{fig:hazard-birth} show roughly constant mortality for the birth cohort 1886--1905 for the entire age range. Mortality is lower for ages 105 and 106 for the birth cohort 1906--1910, but equals that for the earlier birth cohort at ages 107 and above. This reduction in mortality for the later birth cohort implies that plateauing for the entire \Istat{}\ dataset does not start until approximately the age of 108.

\subsection{Graphical diagnostics} \label{sec:qqplot}

\begin{figure}[t!]
	\centering
\includegraphics[width=0.6\linewidth]{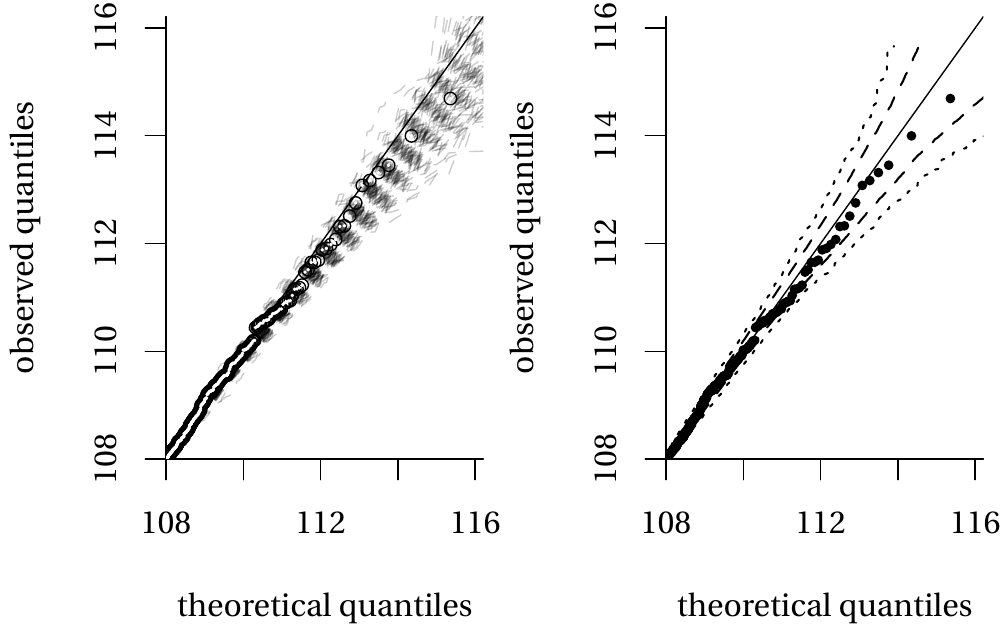}
 \caption{Exponential QQ-plot of the \Istat{}\ data for individuals who lived beyond 108 years and died before 2016, with pointwise (dashed) and simultaneous (dotted) 95\% confidence intervals obtained through simulation (right). The left panel shows trajectories for simulated lifetimes drawn from the exponential model for observed non-censored observations. The circles give the ordered ages at death of non-censored \Istat{}\ lifetimes against the plotting positions.}
 \label{fig:qqplot}
\end{figure}

A quantile-quantile (or QQ-) plot is a standard diagnostic of the fit of a specified distribution to data, but it must be modified to accommodate censoring or truncation. We plot the ordered ages at death, $x_i$, of non-censored \Istat{}\ individuals  against plotting positions from  $\widehat{F}^{-1}\{\widetilde{G}(x)\}$ \citep{Waller/Turnbull:1992}, where in our case $\widehat{F}^{-1}$ is the quantile function of the exponential distribution fitted to ages exceeding 108 years and $\widetilde{G}$ is the Kaplan--Meier estimate of the distribution function, corrected for censoring and truncation \citep{Tsai/Jewell/Wang:1987}. To assess the variability of the plot we simulate new ages at death from the fitted exponential distribution, conditioning on the birth dates, truncation time and censoring indicator; this amounts to simulating new lifetimes from a doubly truncated exponential distribution, since individuals whose death is observed during the sampling frame cannot exceed the age they would reach on 31 December 2015. The left-hand panel of \Cref{fig:qqplot} shows 100 such curves, corresponding to $x_i{}^{(b)}$ on the $y$-axis and $\widehat{F}_{b}{}^{-1}\{\widetilde{G}_{b}(x_i{}^{(b)})\}$; under the parametric bootstrap, both $\widehat{F}$ and $\widetilde{G}$ are re-estimated using the simulated samples. The right-hand panel of \Cref{fig:qqplot} shows approximate 95\% pointwise and simultaneous confidence intervals obtained using a simulation envelope for $\widehat{F}_{b}^{-1}\{\widetilde{G}_{b}(x)\}$ \citep[\S4.2.4]{Davison.Hinkley:1997}. Despite the discrepancy for individuals dying shortly after age 108, an exponential distribution adequately captures the observed mortality.

\subsection{Differences between men and women}
\label{subsect:gendercomparison}

The imbalance in the number of men and women limits our ability to detect any effects of gender on mortality at great age. To illustrate this, we conducted a simulation study based on the \Istat{} data to assess the power of a likelihood ratio test for conditional lifetime exceedances above 108 years, for which we have 40 men and 375 women; the lifetimes of 94 individuals, 15 of them men, are censored. We condition on the sampling frame and the sex of individuals and simulate new life trajectories for both men and women based on an exponential distribution with relative scale differing by a ratio $\lambda>1$, corresponding to lower hazard for women. Thus individuals who were older than 108 in 2009 survive at least beyond that age, but a simulated lifetime that extends beyond 2016 is censored. For each of  10\,000 simulated samples, we computed the likelihood ratio statistic for comparison of fits to men and women separately and a combined fit; the statistic has an asymptotic $\chi^2_1$ distribution. We also considered conditioning only on the birth dates and the beginning of the sampling period to assess whether the right-censoring leads to loss of power, but the difference is negligible. Power of 80\% is achieved if $\lambda\approx1.61$, corresponding to an average survival, $\sigma_e$, of 1.85 years for women and 1.14 years for men, with the corresponding differences for powers 20\% and 50\% being  0.28 years and  0.50 years. The power of the corresponding test for the \IDLold{} data is expected to be similar. 

For the \France{} data a likelihood ratio test strongly rejects the hypothesis of equal hazard for women and men. The ratio of the estimated hazards for men and women in this dataset is approximately 1.61. The combined power of the tests in the \Istat{}{} and \IDLold{} data for detecting this ratio is approximately $0.96$.%$1-0.2^2 = 0.96$. 

%The ratio of the We conclude that appreciably more data would be needed to exclude the possibility that differences in mortality after age 105 due to gender are not detected due to a lack of power. 

%\begin{figure}[!t]
%	\centering
%	\includegraphics[width=0.9\linewidth]{figure/Fig9-power108menvswomen.pdf}
%\caption{Power curve for the likelihood ratio test for difference in exponential hazard between men and women surviving beyond 108 years as a function of the ratio of hazard $\lambda$ between sex. The full line indicates power for the likelihood ratio in which age is censored to its value on January 1st, 2016.}
%\label{fig:powercurve}
%\end{figure}

\subsection{Hazard estimates}%
\label{subsect:nphazard}

To construct a local hazard estimate based on the limiting generalized Pareto distribution, we note that this distribution has reciprocal hazard function  $ r(x) = (\sigma + \gamma x )_{+}$, where $a_{+} = \max(a, 0)$ for real $a$. A more flexible functional form is $r(x) = \{\sigma + \gamma x + g(x)\}_{+}$, where $g(x)$ is a smooth function of $x$ that tends to zero as $x$ increases. For exploratory purposes we take $g(x) = \sum_{k=1}^K \beta_k b_k(x)$, where 
\begin{align*}
b_k(x) = (\kappa_k-x)_{+}^3, \quad  \kappa_1 < \cdots < \kappa_K; 
\end{align*}
here the $\kappa_k$ are fixed knots, and $b_k (x) = 0$ for $x \geq \kappa_k$. The resulting cubic spline function $g(x)$ has two continuous derivatives. This model has $K + 2$ parameters and corresponds to the generalized Pareto distribution for $x > \kappa_K$, but allows $r(x)$ to depart from linearity for $x \leq \kappa_K$. 

Lifetime data are recorded to the nearest day and often there are ties for small $x$, so we use a discrete version of the model, with $x \in \{1, \ldots, x_{\max}\}$ days, where $x_{\max}$ corresponds to 16 years after age 105. We let $h(x) = 1/r(x)$ denote the hazard function, set $H(x) = \sum_{z=1}^x h(z) /365$ for compatibility with the continuous case, and obtain survivor and probability mass functions $\Pr(X > x) = \exp\{-H(x)\}$ and $\Pr(X = x) = h(x) \exp\{-H(x)\}$ for $x \in \{1, \ldots, x_{\max}\}$. 

Let $X$ denote a survival time (days) beyond 105 years.  For each individual the available data are of the form $(s, d, x)$, where $s=0$ if observation of $X$ began at 105 years and, if not, $s > 0$ is the age in days above 105 at which observation of $X$ began, $d = 1$ indicates death, $X = x$, and $d = 0$ indicates right-censoring, $X > x$. The likelihood is then a product of terms of the form 
\begin{align*}
\frac{\Pr(X=x)^d \times \Pr(X>x)^{1-d}}{\Pr(X>s)}, \quad x>s,
\end{align*}
and depends on the parameters $\sigma,\gamma, \beta_1,\ldots, \beta_K$, estimates of which are readily obtained by numerical maximization of the log likelihood function. The resulting fit depends on the knots $\kappa_1,\ldots,\kappa_K$, but to reduce this dependence we generate knots at random, roughly evenly spaced in an interval $(0,x_{\rm max})$, where $x_{\rm max}$ is chosen large enough that $r(x)$ should be linear for $x>x_{\rm max}$, i.e.,  the generalized Pareto model is fitted when $x>x_{\rm max}$.

\begin{figure}[!t]
	\centering
	\includegraphics[width=0.6\linewidth]{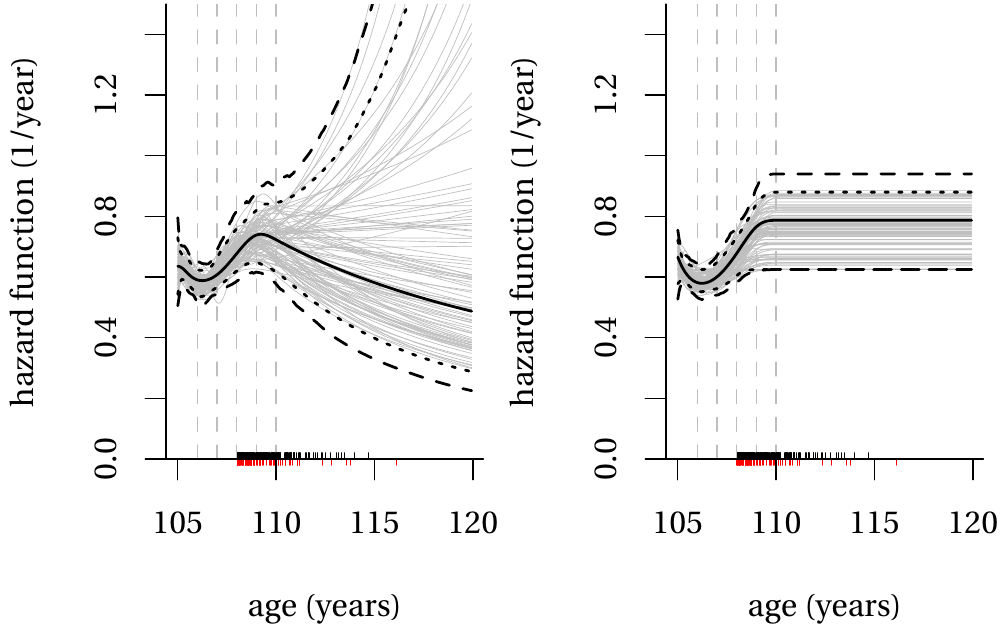}
 \caption{Local hazard estimation for the \Istat{} data using a spline approach with five random knots centered at $ 106, \ldots, 110$ years. Left: original estimate (heavy solid line) with  pointwise (dotted) and overall (dashed) 95\% envelopes obtained from $5000$ bootstrap replicates, of which $100$ are displayed (grey solid lines). The mean positions of the knots are shown by the dashed vertical lines. The black rug shows (jittered) times of deaths, and the red rug shows (jittered) censored survival times; the rugs are suppressed for the lower ages, as there too many points to be informative. The right-hand panel shows the same output for the best-fitting exponential model, for which $\gamma=0$.}
\label{fig:hazard}
\end{figure}

The left-hand panel of \Cref{fig:hazard} shows a local estimate of the hazard function for the \Istat{} data, constrained to have the form of~\eqref{eq:GPhazard} above 110 years. The hazard dips up to 108 years or so, then rises and declines slowly. To assess the significance of this decline we performed a bootstrap analysis \cite{Efron.Tibshirani:1993,Davison.Hinkley:1997}, generating 5000 replicate datasets, the hazard function estimates for 100 of which are shown in the panel. These suggest that the slow decrease after age 110 is not significant, and this is confirmed by the pointwise and overall 95\% confidence bands. The initial dip seems to be a genuine feature, but above 108 years the confidence bands include a wide range of possible functions, including a constant hazard. 
%The right-hand panel shows that the corresponding results with $\gamma=0$ qualitatively agree with the cumulative hazard estimate shown in \Cref{fig:cumulativehazard}, whose initial lower slope agrees with the dip in \Cref{fig:hazard}.

\subsection{Power}
\label{subsect:power}

To assess the statistical power of our procedures, we compute the maximum profile likelihood estimates $\widehat{\sigma}_\gamma$ for the original data with $\gamma $ fixed at the values $-0.25, \ldots, 0$ and simulate new excess lifetimes for each $\gamma$, conditioning on the sampling frame. In the \Istat{} data, for example, we calculate the ages of all individuals on January 1st, 2009, retain the dates at which they reach 108 years and sample new excess lifetimes from the generalized Pareto distribution with parameters $(\gamma, \widehat{\sigma}_\gamma)$, right-censoring any simulated lifetimes beyond the end of 2016.  This ensures that the power calculations are as relevant as possible, not only in terms of the sampling scheme but also in terms of the underlying parameters. For each simulated dataset we compute the directed likelihood ratio statistic $r=\sign(\widehat{\gamma}-\gamma)\{2\ell(\widehat{\gamma},\widehat{\sigma}) - 2\ell(0, \widetilde{\sigma}_e)\}^{1/2}$ and the Wald statistic $w = \widehat{\gamma}/\mathsf{se}(\widehat{\gamma})$ for testing the null hypothesis $ \gamma  = 0$ against the alternative $\gamma < 0$, which corresponds to testing for exponentiality against possible upper bounds to the human lifespan. The asymptotic null distribution of the statistics $r$ and $w$ is standard normal and we can assess the quality of this approximation by calculating the Wald statistics when $\gamma=0$. These simulations show that the estimator of the shape parameter is unbiased and normally distributed, but the distribution of the Wald statistics is left-skewed, leading to inflated Type I error. For the power study, we thus use the simulated null distribution for comparison; tests performed in the paper should be considered liberal, meaning their level would be higher than that claimed.

% For the joint test, we pool the data and compare the fit of a generalized Pareto distribution with a common shape parameter and a different scale parameter $\sigma$ for each dataset with that of exponential model with a different $\sigma$ for each dataset.

We proceed similarly for the endpoint $\iota$.  In order to simulate from generalized Pareto distributions with $\iota$ fixed at a given value, we reparametrise the generalized Pareto log likelihood function in terms of $\iota$ and $\sigma$ and estimate the scale parameters $\widehat{\sigma}_\iota$ for each of the original three datasets with $\iota$ fixed, then use the relation $\iota=  u-\sigma/\gamma$ to obtain the three implied shape parameters $\widehat{\gamma}_\iota$.  We then simulate new datasets with the sampling scheme described above, but using the three sets of parameter pairs $(\widehat{\gamma}_\iota,\widehat{\sigma}_\iota)$. For the joint test of the null hypothesis of an exponential distribution, $\iota=\infty$, against the alternative of a common but finite $\iota$, we reparametrize the likelihood in terms of $\iota$ and allow the three datasets to have different values of $\sigma$.

The left-hand panel of \Cref{fig:powerendpoint} shows how the empirical proportion of rejections for a test of nominal size $5\%$  based on the directed likelihood root statistic $r=-\{2\ell_{\mathsf{gp}}(\widehat{\iota}, \widehat{\boldsymbol{\sigma}}) - 2\ell_{\mathsf{exp}}(\widetilde{\boldsymbol{\sigma}}_e)\}^{1/2}$ varies with $\iota=  u-\sigma/\gamma$ for $\gamma<0$, for the \Istat{}, \France{} and the rest of the \IDLold{} data.  Here  $(\widehat{\iota}, \widehat{\boldsymbol{\sigma}})$ are the maximum likelihood estimates for  the generalized Pareto distribution parametrized in terms of a common finite $\iota$ and three scale parameters and $\widetilde{\boldsymbol{\sigma}}_e$ are the maximum likelihood estimates of the three exponential scale parameters under the null hypothesis.

\subsection{Gompertz model}
\label{subsect:gompertz}
%The Gompertz distribution
%\begin{align*}
%F(x) = 1 - \exp\left\{-(e^{\beta x/\sigma }-1)/\beta\right\}, \quad x>0,\quad \beta,\sigma>0                                                                                            \end{align*}
%has long been used for modelling lifetimes and often provides a good fit to data at lower ages.  
%% In this formulation the variable $Y=(e^{\beta X/\sigma}-1)/\beta$ has a standard exponential distribution. 
%The limit as $\beta\to 0$ can be included in the model, and is the exponential distribution with mean $\sigma$.  

The reciprocal hazard function of the Gompertz distribution $r(x) = \sigma\exp(-\beta x/\sigma)$ 
%always has infinite upper limit to its support, so it is not appropriate for testing whether there is a finite upper limit to lifetimes.  Its hazard function, $\sigma^{-1}\exp(\beta x/\sigma)$, can be finite but increasing for all $x$ ($\beta>0$) or constant ($\beta=0$).   One can check that the limiting distribution for threshold exceedances of Gompertz variables is exponential, and that this limit will be attained rather rapidly.  The reciprocal hazard function 
encodes the speed of convergence to the limiting extreme value distribution \citep{Smith:1987};  even if $\beta>0$, $r'(x) = \beta\exp(-\beta x/\sigma)\to 0$ exponentially fast as $x\to\infty$.  Any improvement in the fit of the Gompertz model for exceedances over some threshold would be shown by evidence that $\beta>0$.

\Cref{tab:gompertz} summarises the fit of this model for various thresholds and the \Istat{} data, without sex or cohort effects.  The hypothesis that the Gompertz distribution ($\beta>0$) reduces to an exponential distribution ($\beta=0$)  is a boundary hypothesis, so the  likelihood ratio statistic to test $\beta=0$ does not have the usual approximate $\chi^2_1$ distribution; its large-sample distribution is a 50:50 mixture of a point mass at 0 and a $\chi^2_1$ distribution, sometimes written as $\frac{1}{2}\chi^2_0+\frac{1}{2}\chi^2_1$ \citep{Self.Liang:1987}.  Barbi et al.~\cite{Barbi:2018} do not notice this, leading them to mis-state the significance of the difference of log-likelihoods in their Table 2 --- the likelihood ratio statistic for testing $\beta=0$ against $\beta>0$ is $w= 2\times 0.292 = 0.584$, and $\pr(\chi^2_1>0.584) \approx 0.44$, which is the significance level quoted in~\cite{Barbi:2018}.  In fact the correct (asymptotic) level would be $\frac{1}{2} \pr(\chi^2_1>0.584) \approx 0.22$.  Here the conclusion does not change, but it might in other contexts.

In general, $p$-values obtained using a parametric bootstrap are likely to be more reliable than asymptotic approximations of the form $\frac{1}{2}\chi^2_0+\frac{1}{2}\chi^2_1$ and are preferable for such comparisons.  In the present case this entails simulating independent datasets like the original data from the boundary exponential distribution (the null hypothesis, $\beta=0$), and estimating the $p$-value by the empirical proportion of these simulated datasets in which the likelihood ratio statistics $w^*$ are no smaller than the original value $w$, i.e., $\widehat{\pr}^*_0(w^*\geq w)$, where the asterisk indicates a parametric bootstrap simulation and the subscript indicates that the simulation is under the null hypothesis; see Chapter~4 of \cite{Davison.Hinkley:1997}.  This approach was used to obtain the $p$-values in \Cref{tab:gompertz}, which show that the exponential model is statistically indistinguishable from the Gompertz model from 108 years onwards, though the Gompertz model fits better at 106 years and below for the \Istat{} data and at 107 years and below for the \France{} data.

Table~\ref{tab:deviancecomp} compares the fits of the Gompertz, generalized Pareto and exponential models to the \Istat{} data, with the baseline taken to be an extended generalized Pareto distribution that encompasses all three other models; the details will be reported elsewhere. The generalized Pareto and Gompertz models fit equally well for all thresholds, since the differences between their respective deviances are minimal.  Both are better than the exponential model below 107 years, but not above, in agreement with \Cref{tab:gompertz}.

\begin{table*}[t]
\centering

%\begingroup
\footnotesize
\begin{tabular}{lrrrrrrrr}
   \toprule
&  threshold & $105$  & $106$  & $107$  & $108$  & $109$  & $110$  \\
 \midrule
\Istat{}&$n_u$ & $3836$ & $1874$ & $946$ & $415$ & $198$ & $88$  \\ 
 & $\sigma$ & $1.67\; (0.05)$ & $1.71\; (0.07)$ & $1.48\; (0.08)$ & $1.47\; (0.12)$ & $1.36\; (0.1)$ & $1.35\; (0.2)$ \\ 
  &$\beta$ & $0.05\; (0.02)$ & $0.09\; (0.04)$ & $0.02\; (0.05)$ & $0.02\; (0.07)$ & $0$ & $0$ \\ 
%   $p$-value (asymptotic) & $0.02$ & $0.01$ & $0.35$ & $0.41$ & $1$ & $1$ & $1$ \\ 
&  $p$-value & $0.02$ & $0.01$ & $0.37$ & $0.45$ & $1$ & $1$  \\ 
   \midrule
\France{}&$n_u$ & $9835$ & $5034$ & $2472$ & $1210$ & $550$ & $240$  \\ 
&  $\sigma$ & $1.71\; (0.03)$ & $1.6\; (0.03)$ & $1.55\; (0.05)$ & $1.43\; (0.06)$  & $1.39\; (0.11)$  \\ 
&  $\beta$ & $0.08\; (0.01)$ & $0.05\; (0.02)$ & $0.06\; (0.03)$ & $0.02\; (0.03)$ & $0$ & $0$  \\ 
%   $p$-value (asymptotic) & $0$ & $0$ & $0.01$ & $0.28$ & $1$ & $1$ & $1$ \\ 
&  $p$-value & $0$ & $0$ & $0.02$ & $0.32$ & $1$ & $1$  \\ 
    \bottomrule
\end{tabular}
\caption{Estimates (standard errors) of Gompertz parameters ($\beta$, $\sigma$) for the Italian \Istat{} data (top) and the \France{} data (bottom) for different thresholds. The first row give the number of threshold exceedances ($n_u$), and the bootstrap $p$-values are for the likelihood ratio test of $\beta=0$ against $\beta>0$. Estimates of  $\beta$ reported as zero are smaller than $10^{-7}$.}
\label{tab:gompertz}
%\endgroup
\end{table*}

% latex table generated in R 4.0.2 by xtable 1.8-4 package
% Fri Aug 14 09:52:32 2020
\begin{table*}[!t]
\centering

\begingroup\footnotesize
\begin{tabular}{rrrrrrr}
  \toprule
  threshold & $105$  & $106$  & $107$  & $108$  & $109$  & $110$\\
 \midrule
% Extended GP & $-8507.42$ & $-4126.41$ & $-1996.29$ & $-880.90$ & $-405.22$ & $-169.74$ \\ 
 Gompertz & $0.01$ & $1.65$ & $2.21$ & $0.22$ & $0.68$ & $1.65$ \\ 
 GP & $0.00$ & $2.17$ & $2.21$ & $0.23$ & $0.56$ & $1.05$ \\ 
exponential & $4.16$ & $8.21$ & $2.36$ & $0.28$ & $0.68$ & $1.65$ \\ 
   \bottomrule
\end{tabular}
\caption{Deviances for comparison of extended generalized Pareto (GP) and the exponential, GP and Gompertz sub-models for different thresholds. The rows show the likelihood ratio statistic, i.e., twice the difference in log-likelihood between the specified model and an encompassing extended generalized Pareto model.} 
\label{tab:deviancecomp}
\endgroup
\end{table*}
\vspace{2cm}
\clearpage

% Bibliography
\bibliographystyle{apalike2}
\bibliography{libraryLongevity}

\begin{thebibliography}{}

\bibitem[Barbi et~al., 2018]{Barbi:2018}
Barbi, E., Lagona, F., Marsili, M., Vaupel, J.~W., \& Wachter, K.~W. (2018).
\newblock {The plateau of human mortality: Demography of longevity pioneers}.
\newblock {\em Science}, 360(6396), 1459--1461.

\bibitem[Davison, 2018]{Davison:2018}
Davison, A.~C. (2018).
\newblock `{T}he life of man, solitary, poore, nasty, brutish, and short':
  {D}iscussion of the paper by {R}ootz{\'e}n and {Z}holud.
\newblock {\em Extremes}, 21(3), 365--372.

\bibitem[Davison \& Hinkley, 1997]{Davison.Hinkley:1997}
Davison, A.~C. \& Hinkley, D.~V. (1997).
\newblock {\em {Bootstrap Methods and Their Application}}.
\newblock Cambridge: Cambridge University Press.

\bibitem[Davison \& Smith, 1990]{davison+s:1990}
Davison, A.~C. \& Smith, R.~L. (1990).
\newblock Models for exceedances over high thresholds (with {D}iscussion).
\newblock {\em Journal of the Royal Statistical Society: Series B (Statistical
  Methodology)}, 52(3), 393--442.

\bibitem[de~Haan \& Ferreira, 2006]{dehaanferreira2006}
de~Haan, L. \& Ferreira, A. (2006).
\newblock {\em Extreme Value Theory: An Introduction}.
\newblock Springer.

\bibitem[Efron \& Tibshirani, 1993]{Efron.Tibshirani:1993}
Efron, B. \& Tibshirani, R.~J. (1993).
\newblock {\em {An Introduction to the Bootstrap}}.
\newblock New York: Chapman \& Hall.

\bibitem[Einmahl et~al., 2019]{einmahl-etal:2019}
Einmahl, J.~J., Einmahl, J. H.~J., \& {de Haan}, L. (2019).
\newblock Limits to human life span through extreme value theory.
\newblock {\em J. Amer. Statist. Assoc.}, 114, 1075--1080.

\bibitem[Gampe, 2010]{gampe2010}
Gampe, J. (2010).
\newblock Human mortality beyond age 110.
\newblock In H. Maier \&  et~al. (Eds.), {\em Supercentenarians}  chapter~13,
  (pp.\ 219--230). Heidelberg: Springer-Verlag.

\bibitem[Gavrilov \& Gavrilova, 2019]{Gavrilov/Gavrilova:2019}
Gavrilov, L.~A. \& Gavrilova, N.~S. (2019).
\newblock Late-life mortality is underestimated because of data errors.
\newblock {\em {PLOS} Biology}, 17(2), 1--7.

\bibitem[Gompertz, 1825]{Gompertz:1825}
Gompertz, B. (1825).
\newblock On the nature of the function expressive of the law of human
  mortality, and on a new mode of determining the value of life contingencies.
\newblock {\em {Philosophical Transactions of the Royal Society of London}},
  115, 513--585.

\bibitem[Guarino, 2018]{Guarino:2018}
Guarino, B. (2018).
\newblock Good news for human life spans --- at age 105, death rates suddenly
  stop going up.
\newblock {\em Washington Post}, (pp.\ June 28).

\bibitem[Hanayama \& Sibuya, 2016]{Hanayama/Sibuya:2016}
Hanayama, N. \& Sibuya, M. (2016).
\newblock Estimating the upper limit of lifetime probability distribution,
  based on data of {J}apanese centenarians.
\newblock {\em J. Gerontol. A Biol. Sci. Med. Sci.}, 71(8), 1014--1021.

\bibitem[Hoad, 2019]{Hoad:2019}
Hoad, P. (2019).
\newblock {`People are caught up in magical thinking': was the oldest woman in
  the world a fraud?}
\newblock {\em The Guardian}, (pp.\ November 30).

\bibitem[ISTAT, 2018]{Istat2018}
ISTAT (2018).
\newblock Italian database on semisupercentenarians.
\newblock See paper by {B}arbi et al. for instructions on how to obtain the
  data.

\bibitem[Oeppen \& Vaupel, 2002]{Oeppenetal:2002}
Oeppen, J. \& Vaupel, J.~W. (2002).
\newblock Broken limits to life expectancy.
\newblock {\em Science}, 296, 1029--1031.

\bibitem[Poulain, 2010]{poulin2010}
Poulain, M. (2010).
\newblock On the age validation of supercentenarians.
\newblock In H. Maier \&  et~al. (Eds.), {\em Supercentenarians}  chapter~1,
  (pp.\ 4--30). Heidelberg: Springer-Verlag.

\bibitem[{R Core Team}, 2020]{R2020}
{R Core Team} (2020).
\newblock {\em R: A Language and Environment for Statistical Computing}.
\newblock R Foundation for Statistical Computing, Vienna, Austria.

\bibitem[Rootz\'{e}n \& Zholud, 2017]{rootzen-zholud:2017}
Rootz\'{e}n, H. \& Zholud, D. (2017).
\newblock Human life is unlimited --- but short (with {D}iscussion).
\newblock {\em Extremes}, 20(4), 713--728.

\bibitem[Rootz\'en \& Zholud, 2018]{RootzenZholud:2018}
Rootz\'en, H. \& Zholud, D. (2018).
\newblock Rejoinder to discussion of the paper ``{H}uman life is unlimited ---
  but short''.
\newblock {\em Extremes}, 21, 415--424.

\bibitem[Saplakoglu, 2018]{Saplakoglu:2018}
Saplakoglu, Y. (2018).
\newblock Have humans reached their limit on life span? {T}hese researchers say
  no.
\newblock {\em Live Science}, (pp.\ June 28).

\bibitem[Scarrott \& MacDonald, 2012]{scarrott/macdonald:2012}
Scarrott, C. \& MacDonald, A. (2012).
\newblock A review of extreme-value threshold estimation and uncertainty
  quantification.
\newblock {\em REVSTAT -- Statistical Journal}, 10, 33--60.

\bibitem[Self \& Liang, 1987]{Self.Liang:1987}
Self, S.~G. \& Liang, K.~Y. (1987).
\newblock {Asymptotic properties of maximum likelihood estimators and
  likelihood ratio tests under nonstandard conditions}.
\newblock {\em Journal of the American Statistical Association}, 82, 605--610.

\bibitem[Smith, 1987]{Smith:1987}
Smith, R.~L. (1987).
\newblock Approximations in extreme value theory.
\newblock Technical Report 205, Center for Stochastic Processes, University of
  North Carolina at Chapel Hill.

\bibitem[Thatcher, 1999]{Thatcher:1999}
Thatcher, A.~R. (1999).
\newblock {The long-term pattern of adult mortality and the highest attained
  age (with Discussion)}.
\newblock {\em Journal of the Royal Statistical Society, Series A}, 162, 5--43.

\bibitem[Tsai et~al., 1987]{Tsai/Jewell/Wang:1987}
Tsai, W.-Y., Jewell, N.~P., \& Wang, M.-C. (1987).
\newblock A note on the product-limit estimator under right censoring and left
  truncation.
\newblock {\em Biometrika}, 74(4), 883--886.

\bibitem[Vijg \& Campisi, 2008]{vijg-campisi:2008}
Vijg, J. \& Campisi, J. (2008).
\newblock Puzzles, promises, and a cure for ageing.
\newblock {\em Nature}, 544, 1065--1071.

\bibitem[Waller \& Turnbull, 1992]{Waller/Turnbull:1992}
Waller, L.~A. \& Turnbull, B.~W. (1992).
\newblock Probability plotting with censored data.
\newblock {\em American Statistician}, 46(1), 5--12.

\end{thebibliography}

\end{document}